\pdfoutput=1

\documentclass[%
 reprint,
 amsmath,amssymb,
 aps,
]{revtex4-2}
\usepackage{algorithm,algpseudocode}
\usepackage{graphicx}
\usepackage{dcolumn}
\usepackage{bm}
\usepackage{comment}
\usepackage[normalem]{ulem}

\usepackage{color,xcolor}
\usepackage{eqnarray,amsmath}
\usepackage{xcolor}
\usepackage{array}
\usepackage{booktabs} 
\usepackage{tabularx}

\newcommand{\pp}{{\bm p}}
\newcommand{\rr}{{\bm r}}

\newcommand{\rhat}{{\hat{\bm r}}}

\renewcommand{\AA}{{\bm A}}

\makeatletter
\def\maketitle{
\@author@finish
\title@column\titleblock@produce
\suppressfloats[t]}
\makeatother

\begin{document}

\preprint{APS/123-QED}

\title{BIP: Boost Invariant Polynomials for Efficient Jet Tagging}

\author{Jose M Munoz}%
\email{munozariasjm@hotmail.com}
\affiliation{EIA University, FTA Group, Antioquia, Colombia.}%

\author{Ilyes Batatia}%
\email{ilyes.batatia@ens-paris-saclay.fr}
\affiliation{Engineering Laboratory, 
  University of Cambridge, 
  Cambridge, CB2 1PZ UK; \\ and 
  ENS Paris-Saclay,
  Université Paris-Saclay, 
  91190 Gif-sur-Yvette, France.}%
 
\author{Christoph Ortner}
\email{ortner@math.ubc.ca}
\affiliation{Department of Mathematics, University of British Columbia, 1984 Mathematics Road, Vancouver, BC, Canada V6T 1Z2.}


\begin{abstract}
Given the vast amounts of data generated by modern particle detectors, computational efficiency is essential for many data-analysis jobs in high energy physics.
We develop a new class of physically interpretable boost invariant polynomial features (BIPs) for jet tagging that achieves such efficiency. We show that, for both supervised and unsupervised tasks, integrating BIPs with conventional classification techniques leads to models achieving high accuracy on jet tagging benchmarks while being orders of magnitudes faster to train and evaluate than contemporary deep learning systems.
\end{abstract}

\maketitle

\section{Introduction}
\label{sec:Introduction}
The study of jets generated at particle colliders is a fundamental tool for understanding subatomic interactions and probing the Standard Model (SM). During experiments at high-energy colliders, a collection of the detected particles is analyzed, recording approximately $10^5$ events per second, from which only a small percentage contain helpful physical information. The detection of events of interest in the myriad of observations has motivated the creation of novel algorithmic approaches to perform a classification given the particles that originated the detected shower, known as \textit{jet tagging}.
The task consists of the classification of cascades of particles generated after the beams collide. The cascade of generated events, known as the jet, comprises a set of particles described by their \textit{four-momentum} ($E, \pp$) and possibly additional features as they are reconstructed in the detectors.
 
The difficulty in the classification arises from the similarity in the structure of detected jets at relativistic energies as measured in the laboratory reference frame. This means that the jets substructure is not fully accessible to the detector, as it would be on the center of mass of the interactions; cf.~Fig~\ref{fig:jets_distiguishable}.
\begin{figure}
    \centering
    \includegraphics[width=0.9\linewidth]{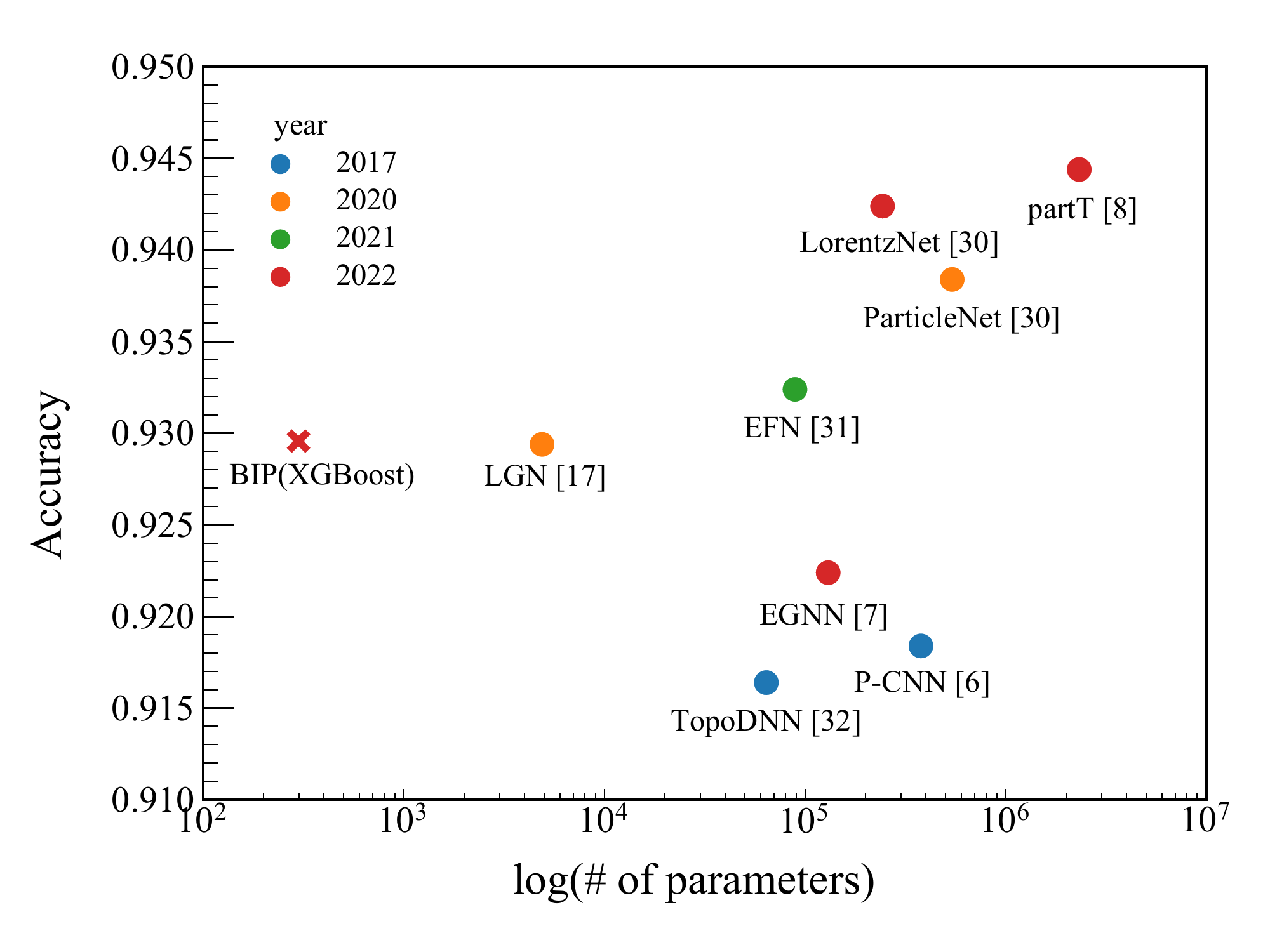}
    \caption{State-of-the-art jet tagging models are compared in terms of the number of parameters.
    While accuracy per parameter is not the key measure to judge the quality of some models, it serves as an illustration of the high performance of BIPs relative to the required computational resources as well as the fact that BIPs provide an entirely new design space for jet tagging models.}
    \label{fig:model_superiority}
\end{figure}

The first approaches to tackle the jet tagging task were clustering algorithms based on features derived from quantum-chromodynamic (QCD) theory~\cite{Salam:2010nqg, Larkoski:2017jix, Dasgupta:2018emf, Farhi:1977sg}. Recently, a range of deep learning algorithms have been proposed, including Convolutional Neural Networks (CNNs)~\cite{Komiske:2016rsd, Macaluso:2018tck} or Graph Neural Networks (GNNs)~\cite{Gong:2022lye, Qu:2022mxj, Shimmin:2021pkm, Ju:2020tbo, Mikuni:2020wpr}. Some efforts have also turned towards the usage of physically inspired features via so-called Energy Flow Polynomials~\cite{Komiske:2017aww, Romero:2021qlf}, or novel QCD-inspired features~\cite{Fedkevych:2022mid, Khosa:2021cyk, Fedkevych:2022mid}. Currently, the most accurate machine learning approaches on jet tagging benchmarks are Lorentz group equivariant message passing networks (LE-MPNNs)~\cite{Erdmann:2018shi, Bogatskiy:2020tje, Gong:2022lye}. This approach exploits the fact that the center of mass is accessible through Lorentz boosts and rotations.
LE-MPNN models are computationally highly demanding as both symmetrization to the full Lorentz group is costly, in addition to a large number of parameters and hence the need for large amounts of training data.

In this letter, we propose a new framework for the jet tagging problem: We construct $N$-body polynomial features that are invariant under (i) permutations of the detected particles in the jet; (ii) boosts {\em in the mean jet direction}; and (iii) rotations around the jet mean axis. By adapting ideas from the Atomic Cluster Expansion (ACE),~\cite{RalfPhysB2019, Dusson22, Kaliu22} we achieve this in a computationally efficient, systematic, and general way. Since the three groups (permutations, boosts, rotations) completely decouple our resulting features are particularly straightforward to derive and implement. We demonstrate the expressiveness of our novel representations by using them as input features for a range of standard classifiers, for both supervised and unsupervised learning. Our emphasis is on simplicity, for example avoiding extensive hyperparameter tuning. Nevertheless, our proposed method achieves excellent accuracy at a computational cost several orders of magnitude lower than state-of-the-art LE-MPNNs, reducing the training time on a large data set to minutes and inference time per jet to tens of microseconds, all on standard CPU hardware and with a small number of parameters (Fig.~\ref{fig:model_superiority}). At the same time, we maintain excellent interpretability and nearly state-of-the-art accuracy in both labeled and unlabeled tasks.

\section{Methodology}\label{sec:methods}

\subsection{Boost-invariant polynomials}
%

\subsubsection{Coordinate transform}
For each (detected) particle $i$ in the jet, experiments or simulations are able to extract their four-momentum, and possibly also additional features $\xi$ (highly application dependent) such as the particle-id, charge, mass, or flavor. A jet can then be understood as a collection of particles $\{ E_i, \pp_i, \xi_i \}_{i=1}^N$.  
The {\em mean direction of the jet $j$} is given by the mean direction of the detected particles, i.e., 
\begin{equation} \label{eq:cylindrical}
    \rr_{\rm jet} = N^{-1} \sum_i \pp_i.
\end{equation}
%
%
We transform the spatial momentum $\pp_i$ to cylindrical coordinates
\[
    \big( p_{\perp,i}, \varphi_i, p_{\|,i} \big)
\]
corresponding to transverse, angular, and parallel components relative to the jet axis $\hat{\rr}_{\rm jet}$. We define the angle via a Householder reflection, which we detail in the SI.

\begin{figure}
    \centering
    \includegraphics[width=0.8\linewidth]{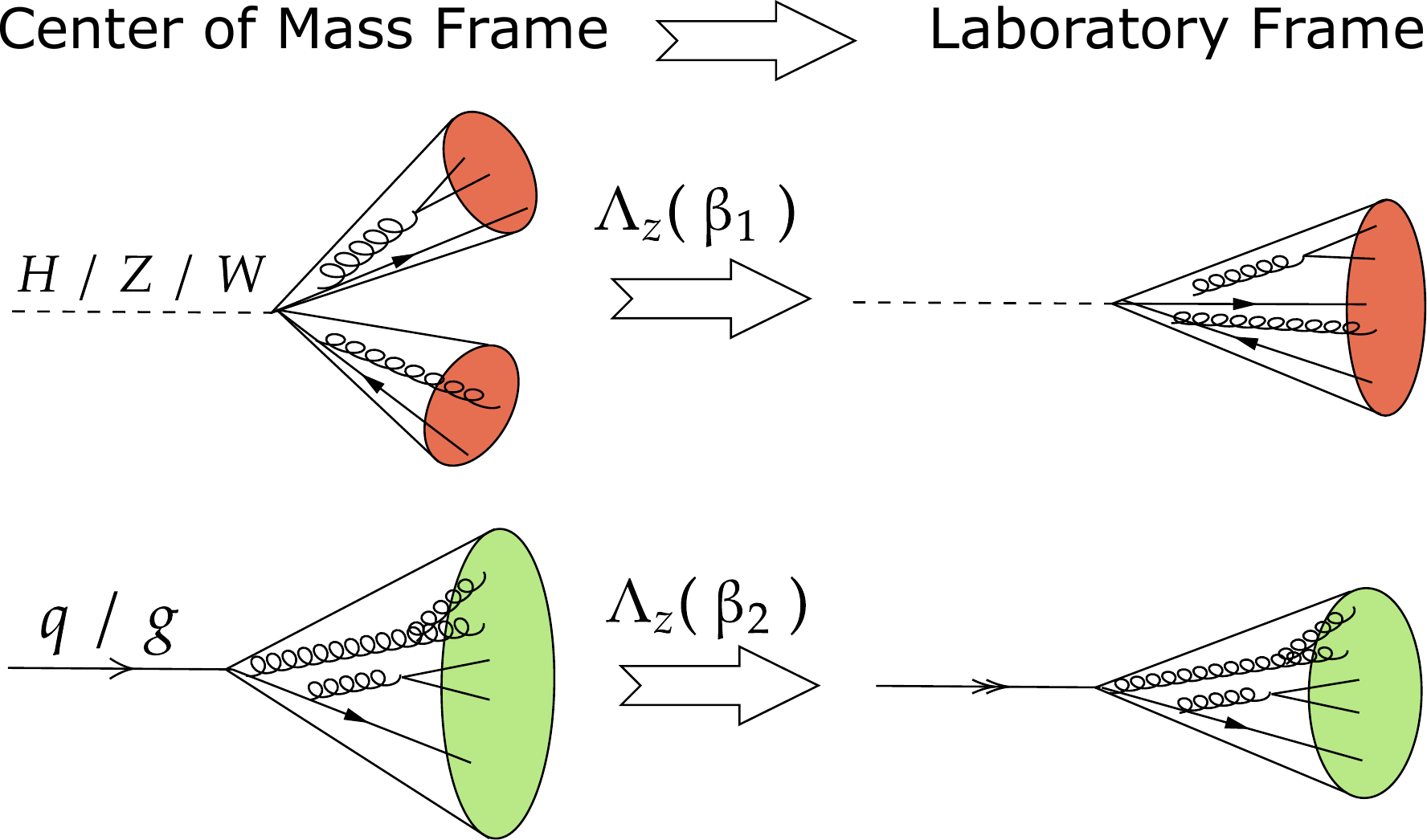}
    \caption{
    Illustration of the transformation of colliding event from the center of mass frame (on the left) to the laboratory frame (on the right). Easily distinguishable events on the left can be similar on the right, making the classification task harder.
    }
    \label{fig:jets_distiguishable}
\end{figure}

Recall that we wish to impose boost invariance in the jet direction which effectively transforms a jet into its center-of-mass frame; cf~Fig.~\ref{fig:jets_distiguishable}.
To give the boost operation a particularly simple form, we introduce the (regularized) rapidity and transverse energy, 
\begin{equation}
    \begin{split}
    y_i &= \frac{1}{2}{\rm log}\left(\frac{\delta_1 + E_i + p_{\|i}}{\delta_1 + E_i -  p_{\|,i}} \right),  \\
    E_{\perp,i} &= \sqrt{m_i^2 + p_{\perp,i}^2 },
    \end{split}
\end{equation}
where $\delta_1 > 0$ is a regularisation parameter to ensure that $y_i$ is well-defined as $E_i -  p_{\|,i} \to 0$. Provided that $\rr_{\rm jet} \neq 0$, the mapping $(E_i, \pp_i) \mapsto (E_{\perp,i}, p_{\perp,i}, \varphi_i, y_i)$ is injective, i.e., it is a genuine coordinate transformation.
In the coordinates $(E_{\perp,i}, p_{\perp,i}, \varphi_i, y_i)$ the effect of a boost $\Lambda_\beta$ in the direction $\rhat_{\rm jet}$ of the jet, and a rotation $R_{\Delta\varphi}$ about the boost axis becomes 
\begin{equation}
    \label{eq:boost-effect}
    \begin{split} 
        & R_{\Delta\varphi} \Lambda_\beta \big(E_{\perp,i}, p_{\perp,i}, \varphi_i, y_i\big) \\
        &= 
    \big(E_{\perp,i}, p_{\perp,i}, \varphi_i + \Delta\varphi, y_i + {\rm tanh}^{-1} \beta_i \big).
    \end{split} 
\end{equation}
Thus, we need only consider the product of two one-dimensional translation groups which will make it particularly straightforward to construct invariant features in a systematic way.

\subsubsection{Many-body polynomial expansion}
The Atomic Cluster Expansion (ACE)~\cite{PhysRevB.99.014104, Dusson22} was proposed as a complete set of polynomial basis functions invariant to rotations and permutations to parameterize a many-body expansion of local interatomic interaction for molecular simulation. 
Our polynomial construction takes heavy inspiration from the ACE expansion, applying analogous techniques {\em globally} rather than locally, and adapting them to the different coordinate systems and symmetry groups that arise in the jet tagging context. 

First, we expand the coordinates of each jet into power sum polynomial type permutation invariant features
\begin{equation}
\label{eq:permutation-basis}
    A_{nlk} = \sum_{i=1}^N Q_n(p_{\perp,i}, E_{\perp,i}, \xi_i) 
        e^{\mathrm{i} l \varphi_i} e^{- \lambda k y_i},
\end{equation}
where we have canonically chosen trigonometric and Morse polynomials (with $\lambda > 0$, though we use $\lambda = 1$ throughout) to embed the angle and rapidity, ensuring the simplest representation of the rotation and boost groups. The features $(p_{\perp,i}, E_{\perp,i}, \xi_i)$ are already invariant, hence we can embed them using a general basis $Q_n$, for which there is considerable design freedom. This freedom makes it possible to account for any additional detected features, $\xi_i$, of particles in a jet. We discuss concrete choices in \S~\ref{sec:Qn}. 

We seek permutation, boost and rotation invariant features of jets. 
All $A_{nkl}$ features are permutation invariant, but only the $A_{n00}$ features are also invariant under rotations and boosts. We can generate a much richer set of permutation, rotation and boost invariant polynomials by forming the product basis
\begin{equation}
    \label{eq:product-basis}
    \begin{split}
    \AA_{\boldsymbol{nlk}} &= \prod_{t=1}^{\nu} A_{n_t l_{t}k_{t}}, \qquad \text{where} \\ 
    \boldsymbol{lkn} &= (n_{1}l_{1}k_{1},\ ...,\ n_{\nu}l_{\nu}k_{\nu}) \quad \text{and } \nu > 0.
    \end{split}
\end{equation}
Only the basis functions satisfying the constraints 
\begin{equation}
\label{eq:filtering}
    \sum_{t} l_{t} = \sum_{t} k_{t} = 0
\end{equation}
which encode, respectively, rotation and boost invariance are retained. The {\em correlation order} $\nu$ indicates how many particles are directly interacting in such a feature. 

With suitable choice of $Q_n$, the $\AA_{\boldsymbol{nlk}}$ features form a complete basis of invariant polynomials (cf. SI), hence any (smooth) property of a jet that satisfies the same invariance can be represented to within arbitrary accuracy as a linear combination, 
\begin{equation} \label{eq:def_bip}
    f\big( \{E_i, \pp_i, \xi_i\}_i \big) = 
    \sum_{\boldsymbol{nlk}} w_{\boldsymbol{nlk}} \AA_{\boldsymbol{nlk}},
\end{equation}
where $w_{\boldsymbol{nlk}}$ are the model parameters (or, weights). 
We call such linear models {\em Boost Invariant Polynomials} (BIPs). The specific selection of feature multi-indices $\boldsymbol{nlk}$ is again application-dependent; we present a simple and general strategy in \S~\ref{sec:sparsefeatures}.

\subsection{Interpretation}
Two intuitive interpretations of the features \eqref{eq:product-basis} are related, respectively, to signal processing and to the many-body expansion. In the context of molecular simulation, analogous connections were explored in detail in \cite{ChemRev2021}. For the sake of a more succinct notation we now identify $v = (n, l, k)$ and $x_i = (E_i, \pp_i, \xi_i)$.

{\it Signal processing interpretation:} Instead of a set of particles, a jet can also be identified with a density
\[
    \rho(x) = \sum_{i = 1}^N \delta(x - x_i),
\]
which can be thought of as a signal. 
Defining the one-particle basis function $\phi_v(x) = Q_n(p_\perp, E_\perp, \xi) e^{i l \varphi} e^{- k y}$ the features $A_{v}$ can be written as projection of the signal onto that basis, 
\[
    A_{v} = \langle \phi_v \,|\, \rho \rangle.
\]
That is, the features $A_v$ {\em represent} the signal $\rho$. Invariant representation  can be obtained by taking the projected $\nu$-correlations, 
\begin{align*}
     \big\langle 
        & \phi_{v_1} \otimes \cdots \otimes \phi_{v_\nu} \,\big|\, 
        \rho \otimes \cdots \otimes \rho \big\rangle \\
        &\qquad = \prod_t \langle \phi_{v_t} \,|\, \rho \rangle 
        = \prod_{t} A_{v_t} = \AA_{\bm v},
\end{align*}
and then averaging them over rotations and boosts. In our current setting this simply results in the constraint in Eq.~\eqref{eq:filtering}.
For this reason we often call the features \eqref{eq:product-basis} symmetry-adapted $\nu$-correlations.

{\em Many-body expansion interpretation:} Let $f$ be a property of a jet that is invariant under permutations, rotations about the jet direction and boosts in the jet direction. Then we can approximate it to within arbitrary accuracy using a many-body expansion, 
\begin{equation} \label{eq:manybody}
    \begin{split}
    f\big( \{ x_i \}_i \big) 
    &= 
    f_0 + \sum_i f_1(x_i)  
     + \sum_{i_1, i_2} f_2\big( x_{i_1}, x_{i_2} \big) \\ 
    & \quad + \dots + \sum_{i_1, \dots, i_{\bar{\nu}}} 
    f_{\bar{\nu}}\big( x_{i_1}, \dots, x_{i_{\bar{\nu}}} \big).
    \end{split}
\end{equation}
Crucially, we include self-interaction in this expansion by allowing the indices $i_1, i_2, \dots$ to be unordered and repeated. Expanding each $f_\nu$ in terms of the tensor-product basis $\phi_{v_1} \otimes \cdots \otimes \phi_{v_\nu}$ and reorganising the summation (see the SI in~\ref{SI} for the details) results exactly in Eq.~\eqref{eq:def_bip} with the $\nu$-correlation features $\AA_{\bm v}$ arising exactly from the expansion of $f_\nu$. Thus, we can alternatively interpret Eq.~\eqref{eq:def_bip} as an efficient linear parametrization of the many-body expansion in Eq.~\eqref{eq:manybody} and the $\nu$-correlation features as natural basis functions for the $\nu$-body term.

\subsection{Jet Tagging with BIPs} \label{bip2tags}

The BIP basis is a {\em complete linear basis} and therefore expressive enough to contain information about the jet's substructure, regardless of the frame observing the interactions. Any classification technique, either linear or nonlinear, can be used to produce a probability score. To that end, we discuss how to select a finite subset of the BIP features, and then explain how we will use them for jet tagging in supervised and non-supervised manners. We emphasize that we will employ no hyper-parameter tuning for the BIP methods that we report.

\subsubsection{The invariant embedding $Q_n$}
\label{sec:Qn} 
There is significant freedom in the design of the embedding $Q_n$ of the invariant features $p_{\perp,i}, E_{\perp,i}$ and $\xi_i$, and for the most challenging data analysis tasks or in the absence of a clear intuition we advocate that it is chosen trainable, e.g. a classical MLP. However, we found that much simpler specifications may often suffice.

We first focus on the case when only the four-momentum is detected (thus ignoring $\xi_i$) as in the top-tagging benchmark; cf. \S~\ref{sec:top_tagg_dataset}. In this case we choose 
\begin{equation}
    \label{eq:def_q}
    Q_n(E_{\perp,i}, p_{\perp,i}) 
    = 
    B_n(\tilde{p}_{\perp,i}) \log(1 + E_{\perp,i}),
\end{equation}
where $B_n$ are the Bessel polynomials applied to the log-transverse momentum 
\begin{equation} \label{eq:transform_p}
    \tilde{p}_{\perp,i} = A \log\Big( { \frac{p_{\perp,i}}{\sum_i p_{\perp,i}}} + \delta_2 \Big) + B.
\end{equation}
Here, $\delta_2$ is another regularisation parameter that ensures $\tilde{p}_{\perp,j}$ remains bounded as $p_{\perp,i} \to 0$ and $A, B$ define an affine transformation to ensure that $\tilde{p}_{\perp,i}$ belongs to the domain of orthogonality of the Bessel polynomials. The logarithmic transformation is suggested by analyzing the distribution of the $p_{\perp,i}$ in the top-tagging dataset. The factor $\log(1 + E_{\perp,i})$ imposes a form of infrared safety (cf. SI) ensuring that particles with low transverse energy do not contribute significantly to the features. We explain in the SI that the resulting embedding $Q_n$ is not theoretically complete as it does not give full flexibility to dependence on $E_{\perp,i}$. We found empirically that providing additional flexibility led to overfitting, and speculate that enough information about the energy of a particle may already be contained in the rapidity variable. 

As an example, how to incorporate additional particle features $\xi_i$ into the embedding $Q_n$ we consider the case when the particle's charge $q_i$ is known. We then set $\xi_i = q_i$ and incorporate it through a one-hot embedding,
\[
    Q_{nq}(E_{\perp,i}, p_{\perp,i}, q_i) 
    = 
    B_n(\tilde{p}_{\perp,i}) \log(1 + E_{\perp,i}) \delta_{q, q_i},
\]
also changing $n$ to a multi-index $(n, q)$.

\subsubsection{A priori sparse feature selection}
\label{sec:sparsefeatures}
The infinite set of possible BIP features in \eqref{eq:product-basis} is indexed by a high-dimensional multi-index. We, use a common sparse grid technique to select which features to employ for classification tasks. First, we fix an upper bound $\bar\nu$ on the correlation order, which is a measure of how strongly correlated groups of particles are. Secondly, we specify a {\em level} $\Gamma$, which is primarily an approximation theoretic parameter. We now select all features $\boldsymbol{nlk}$ satisfying 
\begin{equation}
    \label{eq:totaldegree}
    \sum_{t = 1}^\nu |l_t| + |k_t| + n_t \leq \Gamma
    \quad \text{and} \quad \nu \leq \bar\nu.
\end{equation}
In the limit as $\bar\nu \to \infty$ and $\Gamma \to \infty$, we recover {\em all} possible features, and in this limit our model becomes universal.
We label the resulting model a ${\rm BIP}(\bar\nu, \Gamma, {\rm method})$ where ``method'' stands for the technique we use to create a classifier from the features and which we detail next.

\subsubsection{Supervised learning}
To classify jets, we arrange the BIP basis as a vector and use it as input into standard classification schemes. For each data point, the label contains the expected classification to a selection of standard classifiers without any modification from the default values in the published implementation in \cite{scikit-learn}. Our method of choice is ensemble learning (Gradient Boosting), but we also test BIP features in conjunction with neural networks (a standard multi-layer perceptron), and linear classifiers (with logistic regression and support vector machine). We observe that the accuracy does not change significantly between all of these classifiers. A further examination of this phenomenon and a description of the models is given in the SI \S~\ref{sec:models}.

\subsubsection{Unsupervised learning}

Following proposals in~\cite{Komiske:2022vxg, Alvarez:2021zje}, we study unsupervised learning using the BIP method. Unsupervised learning is of interest for several reasons: It can identify deviations between observed and simulated data and could therefore be employed to detect physics beyond the standard model. It avoids bias towards the detector's nuisance parameters, which heavily increase systematic uncertainties. Finally, it helps reduce the need for training data in order to perform phenomenological analyses.

Since higher-dimensional spaces tend to make distance metrics asymptotically indistinguishable~\cite{thrun2021exploitation}, we first perform a Uniform Manifold Approximation and Projection (UMAP) developed in \cite{mcinnes2018umap} for dimensionality reduction inspired by the t-SNE approach used in \cite{Komiske:2019fks}. 
%
%
After the embedding has been projected, we show the expressiveness of this ultra-compact feature set by training a Gaussian Mixture Model and a k-means clustering algorithm. Further details are given in the SI, \S~\ref{sec:models}.

\section{Results}

\subsection{Top tagging benchmark}
\label{sec:benchmark_1} 
\label{sec:top_tagg_dataset}
We show a selection of BIP model results for jet tagging, comparing them to previous approaches. 
We compare the classification performance of a range of models via the {\em accuracy} and {\em Area Under the Curve} (AUC) measures and contrast this against the number of parameters employed. 
A wider range of BIP results for different out-of-the-box models is given in the SI in ~\ref{SI}. 

The {\em top tagging dataset} was proposed by~\cite{Kasieczka:2019dbj} performing a Delphes simulation with cuts on $R = 0.8$ and $\eta\leq 2$. It consists of 1.2M training, 400k validation, and 400k test data points. Each data point represents a jet whose origin is a top quark, a light quark, or a gluon. Each jet is an array of $N$ detected particles in four-momentum coordinates $(E, \pp)$. On average, there are 30 particles per jet but a maximum of 200 constituents.

Table~\ref{top-tagging-table} shows that our BIP embedding, together with a linear classifier, can reach excellent accuracy using several orders of magnitude fewer parameters resulting in a total training time under 50 seconds.
The unsupervised setting shows BIP's full expressivity, enabling the Gaussian Mixture classifier to reach excellent accuracy with only 5 parameters.

\begin{table}

\label{top-tagging-table}
\vskip 0.15in
\begin{center}
\begin{small}
\begin{tabular}{lccr}
\toprule
\midrule
Architecture & \#Param & Accuracy & AUC \\
\midrule \midrule
\textbf{*partT (2022)}\cite{Qu:2022mxj}    & 2.14M & 0.944 & 0.988  \\
\textbf{EGNN (2022)}\cite{Gong:2022lye}    & 120k & 0.922 & 0.970  \\
\textbf{PCT (2021)}\cite{Qu:2019gqs}  & {139.3k}  & 0.940 & 0.986 \\
\textbf{EFN (2021)}\cite{Komiske:2018cqr}    &  82k & 0.927 & 0.979 \\
\textbf{ParticleNet (2020)}\cite{Qu:2019gqs} & 498k & 0.938 & 0.985  \\
\textbf{LGN (2020)}\cite{Bogatskiy:2020tje}   &  {4.5k} & 0.929 & 0.964 \\
\textbf{P-CNN (2017)}\cite{Macaluso:2018tck}   & 348k & 0.918 & 0.980  \\
\textbf{TopoDNN (2017)}\cite{Pearkes:2017hku}  &  59k & 0.916 & 0.972\\
\midrule

Supervised &  &  & \\
\textbf{BIP}(3, 6, MLP) &  {4k} & 0.931 & 0.981 \\
\textbf{BIP}(3, 6, XGBoost) &  {300} & 0.929 & 0.978 \\
\textbf{BIP}(3, 6, LogReg) &  {300} & 0.927 & 0.977 \\
\textbf{BIP}(3, 6, SVM) &  {300} & 0.927 & 0.976 \\[1mm]
Unsupervised &  &  & \\
\textbf{BIP}(3, 6, UMAP+GMM) &  {5} &  0.864 & 0.898 \\
\textbf{BIP}(3, 6, UMAP+KNN) &  {2} &  0.845 &  - \\
\bottomrule
\end{tabular}
\end{small}
\end{center}
\vskip -0.1in
\caption{Performance comparison between BIP classifiers and a range of other classifiers taken from \cite{Kasieczka:2019dbj, Qu:2019gqs}. We report the results as BIP($\bar\nu$, $\Gamma$, architecture); in all cases, 300 BIP features are used. ${\bm (*)}$ The {\tt partT} model was pre-trained on a much larger dataset and then fine-tuned to the top-tagging dataset, which accounts for the improved performance in this comparison.}
\end{table}

\subsection{Versatility and Efficiency}
Our model is highly computationally efficient; to demonstrate this we performed a performance benchmark on an AMD EPYC-Rome Processor using a fully serial framework and with no GPU usage. The computational pipeline involves custom data processing and internal transformations explained in Sec. \ref{sec:methods}. This stage of the computation takes $2.251\ \mu\text{s} \pm  5.928\  \mu\text{s}$ per jet.

\begin{figure}
    \centering
    \includegraphics[width=\linewidth]{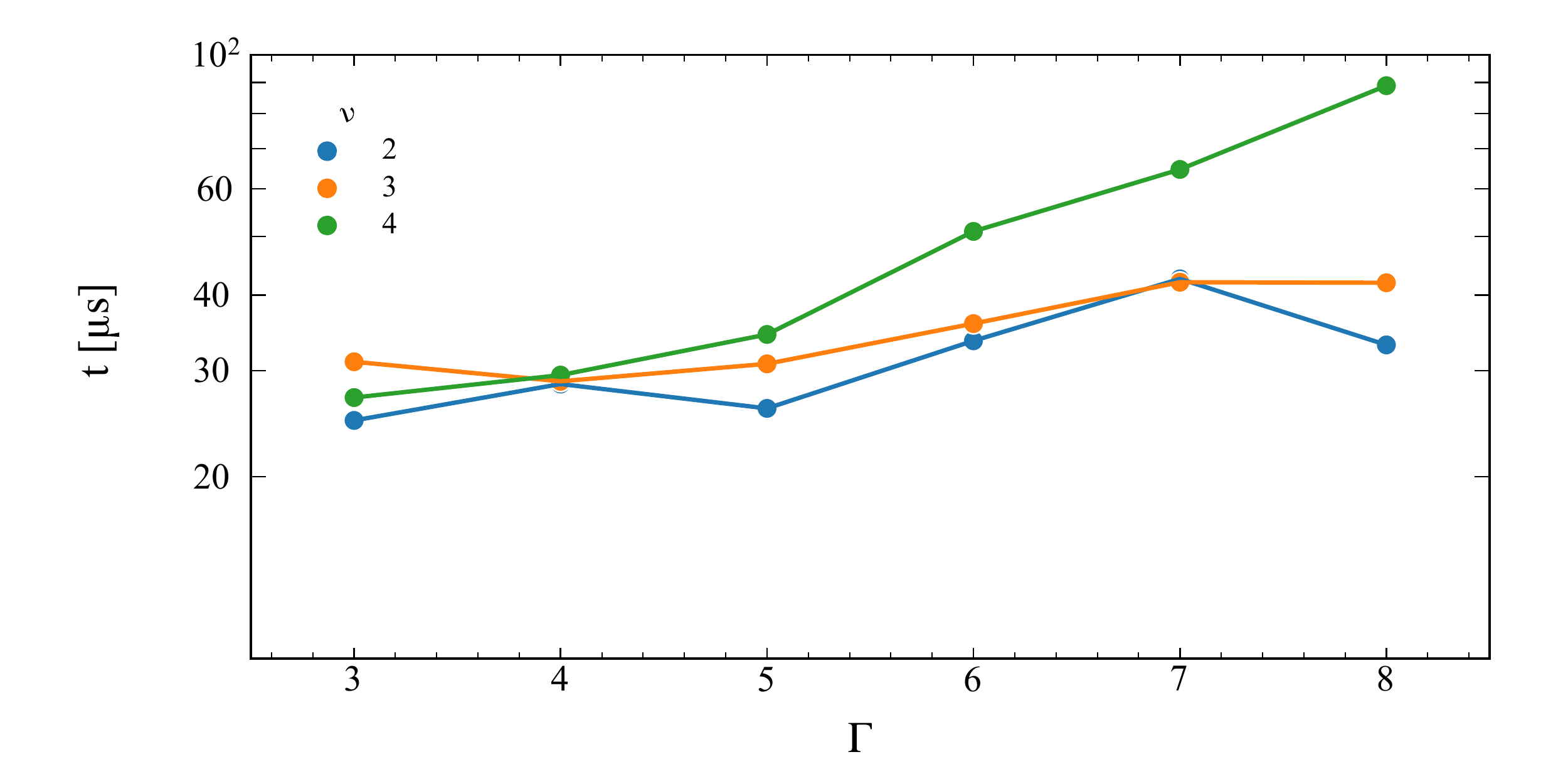}
    \vspace{-3mm}
    \caption{Performance of the construction of the BIP features at different correlation order $\nu$ and level $\Gamma$, the standard deviation is calculated by performing the transformation for all the training sets in a sequential mode with no parallelization. 
    }
    \label{fig:speed_order}
\end{figure}

The density projections $A_{nlk}$, as well as the $\nu$-correlations \eqref{eq:product-basis}, are both fast to evaluate, even without the optimal algorithm proposed in \cite{Kaliu22}, meaning that no usage of a GPU is required to obtain excellent performance, as shown in Fig.~\ref{fig:speed_order}. 

\begin{figure}
    \centering
    \includegraphics[width=1.0\linewidth]{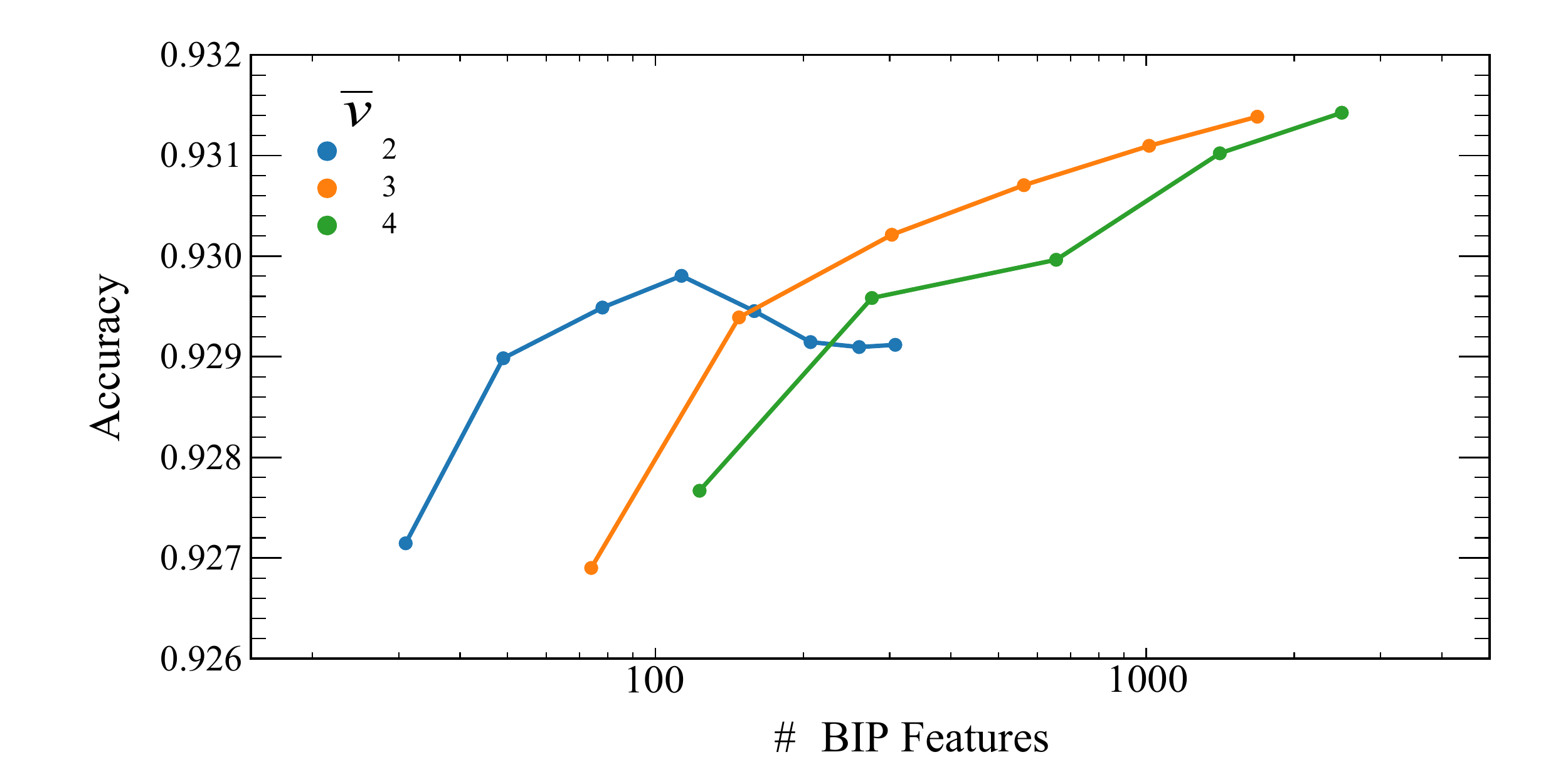}
    \vspace{-3mm}
    \caption{The accuracy as a function of the sparsification parameters where increasing the number of features is obtained via increasing $\Gamma$. See the Text for more details.}
    \label{fig:level_acc}
\end{figure}

The performance depends slightly on the correlation order $\bar\nu$ and level $\Gamma$ parameters of the BIP framework, which gives a variable basis size and allows us to trade efficiency against the computational cost (c.f Fig.~\ref{fig:level_acc}).
This means that as a general rule, the accuracy vs computational cost trade-off can be tuned according to the application. For instance, for trigger-level applications or prototyping for model searches, computational performance may be the main requirement, while the trade-off might be slanted toward accuracy when final statistical analyses are conducted.

\section{Conclusions and Outlook}
We introduced the boost-invariant polynomials (BIP) model, a systematic, interpretable, and highly efficient jet tagging architecture employing polynomial features that are invariant under permutations, rotations about the jet direction, and boosts in the jet direction. The framework draws many ideas from the Atomic Cluster Expansion Model~\cite{RalfPhysB2019,Dusson22} to achieve its generality and efficiency. 
While our approach does share concepts with energy flow polynomials~\cite{Komiske:2017aww} due to the fact that both are building polynomial features, our architecture is entirely different, employing different coordinate systems, invariances, and without requirement for a pairwise metrics.

We speculate that due to the simplicity of its architecture it might be easily implemented using Field-Programmable Gate Arrays (FPGAs) as dedicated hardware for employing our framework in an experimental setting. In addition, our construction is highly versatile and enables us to easily incorporate additional measured particle properties (e.g., charge, spin, flavour).

Despite the simplicity of our approach, we achieve close to state-of-the-art accuracy while gaining several orders of magnitude of speedup in the training and inference stages. It is maybe particularly remarkable that we achieved those results without employing any hyper-parameter tuning. 


The ACE framework was recently extended as a general framework for equivariant message passing \cite{DesignBatatia2022, MLAYERBochkarev2022} and has proven highly successful for modeling inter-atomic interactions~\cite{Kovacs2021,batatia2022mace} typically outperforming other approaches despite employing much shallower architectures. This work suggests that the BIP model could also be extended to a geometric deep-learning framework which would naturally lead to automated discovery of the embedding $Q_n$, and generally open up further model tuning possibilities that will likely further improve the already excellent accuracy we obtain with BIP models. 

\begin{acknowledgments}
We thank Guillermo Palacio, Jesse Thaler, Sam McDermott, and Jan Offermann for their comments and suggestions on an earlier version of this manuscript. 

This work was initialized while JMM visited UBC funded by a MITACs Globalink internship. CO was supported by NSERC [IDGR019381].
\end{acknowledgments}



\nocite{*}

\bibliography{main}

\clearpage

\title{Supplementary Information \\[2mm] BIP: Boost Invariant Polynomials for Efficient Jet Tagging}

\maketitle

\label{SI}

\setcounter{page}{1}
\setcounter{section}{0}
\setcounter{equation}{0}

\subsection{Miscellaneous}

\subsubsection{Lorentz Group}
The mass and any other tensor product must be invariant under the Lorentz Group denoted by $SO(1,3)$. To be precise, the actual group of interest in particle physics is the {\em ortochronus} Lorentz group, $SO^+(1,3)$, since the time direction of the space-time coordinate may not be negative. However, this distinction is unimportant for our purposes. For any  element $\Lambda \in SO(1,3)$, tensors transform in such a way that the Minkowski metric $\eta = \text{diag}(1,-1,-1,-1)$ is preserved:
\begin{equation}
    \Lambda ^{-1} = \eta \Lambda^T\eta^{-1}
\end{equation}
The group consists of rotations about the three angles and boosts on an arbitrary axis $\hat r$, that can parameterized by $\beta_{\hat{r}} \equiv \|v_{\text{boost}}\|$, giving the Lorentz factor with Natural Units ($c=1$). If a particle with four-momentum $P = (E, p_x, p_y, p_z)$ is boosted along the $z$ axis, the transformation reads:
\begin{equation}
    \Lambda_{\beta_{\hat k}}(P) = P' = 
    \Big(\frac{(E-\beta\cdot p_z)}{\sqrt{1+\beta^2}},
    p_x,
    p_y,
    \frac{(p_z - \beta\cdot E))\big)}{\sqrt{1+\beta^2}}\Big).
\end{equation}
From this expression, one can readily obtain the action of a boost in our transformed coordinates as stated in \eqref{eq:boost-effect}. 

\subsubsection{Householder Transformation}
After the particles are detected with their respective four-momentum components, and the mean jet direction $\rr_{\rm jet}$ is determined we perform a Householder transformation~\cite{goodall199313} to obtain the angular component of the cylindrical coordinate system \eqref{eq:cylindrical}. The procedure is as follows: we compute the mean direction of the spatial momentum of the detected particles $\rr_{\rm jet} = N^{-1} \sum_i \pp_i$. The Householder transformation is then a reflection represented by the matrix 
\[
    H = p - 2 {\bm u} {\bm u}^T
\]
where ${\bm u}$ is chosen such that it lies in the plane spanned by $\rr_{\rm jet}$ and ${\bm e}_3$ and such that $H \rr_{\rm jet} = {\bm e}_3$. This results in 
\[ 
    {\bm u} = {\bm w}/\|{\bm w}\|
    \quad \text{where} \quad 
    {\bm w} =  \rr_{\rm jet} - (\rr_{\rm jet} \cdot {\bm e}_3) {\bm e}_3.
\]    


\subsection{Completeness of the BIP Model}
In this section, we make precise our claim that the linear BIP model is ``complete''. This is the most stringent requirement on a set of features and implies in particular also completeness (universality) of nonlinear models. Our arguments in this section should not be considered mathematically rigorous proof, but they constitute an outline that clarifies the required assumptions and from which a fully rigorous proof can be readily constructed.

\subsubsection{Smooth many-body expansion} 
The starting point of our BIP model is the many-body expansion \eqref{eq:manybody}. Although it can in some few cases be justified rigorously, we have not pursued this and require this as an {\em assumption}: we assume in the following that a property $f_{\rm ref}$ of jets $\{x_i\}_i$, where $x_i = (E_i, \pp_i, \xi_i)$, we wish to represent can be approximated to within arbitrary accuracy by a many-body expansion  \eqref{eq:manybody} with components $f_n$ that are smooth.

By ``approximated`` to within arbitrary accuracy we mean that there exist such components $f_0, \dots, f_{\bar\nu}$ such that the distance between the resulting model $f$ and the target $f_{\rm ref}$, 
\[
    {\rm dist}(f, f_{\rm ref}) := \sup \big| f(\{x_i\}_i) - f_{\rm ref}(\{x_i\}_i) \big|
\]
can be made arbitrarily small. Here, the supremum is taken over all jets $\{x_i\}$ with particles $x_i$ taken from some specified {\em bounded domain} $x_i \in \Omega$. (An extension to unbounded domains is possible but more subtle.)

\subsubsection{Requirements for completeness}
Completeness of $\nu$-correlations in the setting of interatomic potentials or force fields was established in \cite{Dusson22} and further clarified and generalized in \cite{BachDusOrt21}. In those references, it is shown that if the {\em one-particle basis}, 
\begin{align*}
    \phi_v(x_i) &= \phi_{nlk}(p_{\perp,i}, E_{\perp,i}, \xi_i, \varphi_i, y_i) \\ 
    &= Q_n(p_{\perp,i}, E_{\perp,i}, \xi_i) 
        e^{\mathrm{i} l \varphi_i} e^{- \lambda k y_i}
\end{align*}
is complete, then completeness (in the sense of a complete basis) of the $\nu$-correlations/product-basis \eqref{eq:product-basis} follows. 

Our general definition of the one-particle basis is as theoretic arguments imply that, if $Q_n$ is complete, then so is $\phi_{nlk}$. Here we use the fact that $\bar{\Omega}$ is bounded which means that it can be enclosed in a  bounded tensor product domain. Thus, as long as $Q_n$ represents a complete linear basis for smooth functions of $(p_{\perp,i}, E_{\perp,i}, \xi_i)$ it follows that the resulting BIP model is complete as well.  

So far, we have ignored rotation and boost invariance in this discussion. But suppose now, that the target property $f_{\rm ref}$ satisfies those invariances. Then it can be readily seen  that symmetrizing $f$ results in a smaller error: let $G = \mathbb{T} \otimes \mathbb{R}$ where the torus $\mathbb{T}$ represents rotations (translations of $\varphi$) while $\mathbb{R}$ represents boost along the jet axis (translations of $y$). If we define the symmetrization operation
\begin{align*}
    g^{\rm sym}(\{x_i\}_i) &= 
    \lim_{Y \to\infty} \frac{1}{4\pi Y} 
    \int_{-\pi}^\pi d \Delta\varphi \int_{-Y}^Y d\Delta y  \\ 
    &\qquad 
    g\big( \{ p_{\perp,i}, E_{\perp,i}, \xi_i, \varphi_i + \Delta\varphi, y_i + \Delta y \}_i \big), 
\end{align*}
then a direct calculation, only slightly adapting the argument from \cite{Dusson22} shows that 
\[
    {\rm dist}(f^{\rm sym}, f_{\rm ref}) \leq 
    {\rm dist}(f, f_{\rm ref}).
\]
Since $f$ is a linear model, the symmetrization operation can be directly applied to the product basis features ($\nu$-correlations): 
\[
    f^{\rm sym}(\{x_i\}_i)
    = 
    \sum_{\bm v} w_{\bm v} {\bm A}_{\bm v}^{\rm sym}.
\]
Finally, another straightforward direct calculation, using the fact that $Q_n$ is already invariant, shows that 
\[
    {\bm A}_{\bm v}^{\rm sym}
    = 
    \begin{cases}
        {\bm A}_{\bm v}, & \text{if } \sum_t l_t = \sum_t k_t = 0, \\ 
        0, & \text{otherwise.} 
    \end{cases}
\]
This completes the proof of completeness of the general BIP model \eqref{eq:def_bip}.

\subsubsection{Deviations from completeness in our top-tagging model}
Our top-tagging model described in the main text fails to satisfy the completeness requirements in a few minor ways, which we explain next. To begin, we first construct a provably complete top-tagging model: to that end, we simply specify two (e.g., polynomial) bases/embeddings $P_{n_p}^p(\tilde{p}_{\perp,i})$ and $P^e_{n_e}(\tilde{E}_{\perp, i})$ where $\tilde{p}, \tilde{E}$ are suitably transformed variables and take their tensor product to obtain the invariant embedding 
\[
    Q_{n_p,n_E}(p_{\perp,i}, E_{\perp,i}) = P_{n_p}^p(p_{\perp,i}) P_{n_e}^e(E_{\perp,i}).
\]
With this choice, our top-tagging model would indeed be a complete model. In our actual model, we made two modifications: first, we only took a single feature for the transverse energy, $\log(1 + E_{\perp,i})$. This choice significantly reduced overfitting and therefore improved test accuracy. Secondly, our transformation $p_{\perp, i} \mapsto \tilde{p}_{\perp,i}$ defined in \eqref{eq:transform_p} is not a genuine coordinate transformation as it is invariant under joint rescaling the transverse momentum variables, that is even introduces an additional invariance. This rescaling is a common technique in jet-tagging~\cite{Qu:2022mxj, Qu:2019gqs} motivated by the idea that the distribution of transverse momentum rather than transverse momentum itself is the quantity accessible in typical measurements. 



\subsubsection{IRC Safety}
Since observables should be computable in the QCD regime, this implies that tagging models should be robust to uncertainties in modeling the showers. To this end, one commonly aims to have models that are Infrared/Collinear (IRC) safe~\cite{Tkachov_1997}: if the descriptor of a jet is a function $f(P_1, \dots, P_N)$ of the detected four-momenta $P_i$, these are defined as follows: 
\begin{itemize} 
    \item The model $f$ is called infrared safe if it is invariant under the addition of a new constituent with zero energy; that is, 
    \begin{align*}
    &f(P_1, \dots,  P_N)  
    = f(P_1, \dots, P_N, P_{N+1}),
    \end{align*}
    whenever $P_{N+1} = (0, \pp_{N+1})$.
    
    \item A model is called {\it collinear safe} if it is invariant under substituting one of the particles with a group of collinear ones bearing the same total four-momentum. 
\end{itemize}

The BIP framework can easily incorporate infrared safety, simply through the requirement that $Q_n(p_{\perp,i}, 0, \xi_i) = 0$. In our top tagging model, we explicitly enforced this requirement; cf. \eqref{eq:def_q} in the main text. It is shown in \cite{Dusson22} that the completeness of the BIP model remains true when such ``boundary conditions'' are imposed.


The collinear safety mechanism appears to require that the $A_{nlk}$ features are linear in the particle inputs, which is a very stringent requirement that is at odds with the completeness of the model. A closer look suggests that in fact one may need to project the BIP features onto a non-trivial manifold; integrating this constraint into the BIP model is likely more challenging. 



\subsection{Models}\label{sec:models}

Below we give a short description of the out-of-the-box classifiers used.

\subsubsection{Supervised approach}
\label{sec:classificer-descriptions}
\begin{itemize}
    \item \textbf{XGBoost}: We make use of the XGBoost implementation of the Extreme Gradient Boosting developed in~\cite{chen2015xgboost}. The algorithm works b creating and selecting a combination of weak learners in which each step of the optimization.
    \item \textbf{Logistic Regression}: We use the regularized Ridge LogisticRegressor implemented in \cite{scikit-learn}.
    \item \textbf{SVM}: Support Vector Machines (SVM) are one of the most common non-probabilistic classifiers, which works by fitting a decision boundary on a custom kernel. We use the Stochastic Gradient Descent regularized SVM implemented in \cite{scikit-learn}.
    \item \textbf{MLP}: Multi-Layer Perceptrons are the most basic implementation of a Fully-Connected Feed-Forward Neural Network (FCNN). This simple model is implemented by default in \cite{scikit-learn} with 2 hidden layers and a ReLU activation function. 
\end{itemize}

\subsubsection{Unsupervised approach}
\label{sec:unsupervised-appendix}
\begin{itemize}
    \item \textbf{UMAP - GMM}: We present our experiments using also no-supervised learning. First, we reduce the dimensionality of the UMAP to a 2-dimensional manifold. Consequently, we use the implementation in \cite{scikit-learn} of the Expectation Maximization (EM) fitting of a Gaussian Mixture Model with full-covariance and two classes. Notice that the reported accuracy only needs the test dataset, first to transform the BIP embedding, then to fit the GMM, and finally to compute the score. However, we performed the same process for the validation and training datasets independently obtaining similar accuracy.
    \item \textbf{UMAP - KMeans}: Using the same UMAP setup described above, we run the k-Nearest Neighbors (KMeans) clustering algorithm implemented in \cite{scikit-learn} with 2 components and the default parameters.
\end{itemize}
$$$$

\subsection{Key differences with the EFP}
Given there are connections between the BIP framework and Energy Flow Polynomials \cite{Komiske:2017aww}, we briefly sketch out some distinctions:
\begin{itemize}
    \item In general the construction made in EFP requires the introduction of a pairwise distance metric between events. On the other hand, the creation of the embedding presented in Eq.~\ref{eq:permutation-basis} evaluates particle features in an agnostic manner.
    \item The invariance to boost in the beam axis as obtained in the EFP, and as well in most of the standard DL models is achieved when considering the $\Delta y$ for the distance metric in the rapidity-angle plane. By contrast, we build our framework using as an axis the mean momentum of the detected particles, and the direction in which the invariance is achieved. Using this approach we aim to obtain insights into the substructure of the reconstructed center of mass.
    \item Our explicit sparsification of the BIP embedding allows us to obtain a fully interpretable basis without the need to perform any fitting on the coefficients. This fact allows us to obtain state-of-the-art results in an unsupervised approach.
    \item We introduce directly in \eqref{eq:permutation-basis} the possibility to encode further the information gathered per particle. This means that it becomes trivial to add features such as spin and charge of the particles without the need to re-embed the basis as done in the Energy Flow Network approach \cite{Komiske:2018cqr}.
    \item The high computational performance of the embedding generation does not depend on any heuristic to approximate the solution of an NP-hard problem as it is done \cite{Komiske:2017aww} but is a simple consequence of the BIP construction~\cite{Kaliu22}. As a matter of fact, in our experiments we did not yet employ the quasi-optimal evaluation algorithm proposed in \cite{Kaliu22}.
\end{itemize}

\clearpage

\end{document}